\documentclass[runningheads]{svmult}

\usepackage{makeidx}   
\usepackage{graphicx}  
\usepackage{subeqnar}  
\usepackage{multicol}  
\usepackage{physprbb}  
\makeindex             

\begin{document}
\title*{An Intermediate Redshift Supernova Search at\protect\newline 
ESO: Reduction Tools and Efficiency Tests}
\toctitle{An Intermediate Redshift Supernova Search
\protect\newline at ESO: Reduction Tools and Efficiency Tests}
%
%
\titlerunning{An Intermediate Redshift Supernova Search}
%
\author{Marco~Riello\inst{1,2}
\and Giuseppe~Altavilla\inst{1,2}
\and Enrico~Cappellaro\inst{3}
\and Stefano~Benetti\inst{1}
\and Andrea~Pastorello\inst{1,2}
\and Ferdinando~Patat\inst{4}
\and Marco~Prevedello\inst{1}
\and Massimo~Turatto\inst{1}
\and Luca~Zampieri\inst{1}}
\authorrunning{Marco~Riello~et~al.}
%
%
\institute{INAF - Astronomical Observatory of Padova, Vicolo dell'Osservatorio 5,\\
           I-35122 Padova, Italy
\and Department of Astronomy, University of Padova,
     Vicolo dell'Osservatorio 2,\\ I-35122 Padova, Italy
\and INAF - Astronomical Observatory of Capodimonte,  Via Moiariello 16,\\
     I-80131 Napoli, Italy
\and European Southern Observatory, Karl-Schwarzschild-Str. 2,\\
     D-85748 Garching bei M\"{u}nchen, Germany }

\maketitle              

\begin{abstract}
\index{abstract}
We present the reduction and archiving tools developed for our search for
supernovae at intermediate redshifts at ESO as well as the efficiency tests
performed. The data reduction recipes developed for the SN candidates 
selection are described. All the variable sources detected are stored using
a MySQL database which enables the identification of previously detected 
variable sources during past observational runs. Finally, experiments 
performed with artificial stars have shown that seeing plays a crucial role
for the limiting magnitude of detection. Crucial is also the detection
threshold used by Sextractor.
\end{abstract}

\section{Introduction}
In the last couples of years we have carried out a SN search with
the main scientific aim of determining the SN rate at intermediate redshift
$(0.2\leq z\leq0.5)$ (for details see the
contribution by Altavilla et al. in this volume). For the detection of SN 
candidates we use the Wide Field Imager at ESO/MPE 2.2m telescope (La Silla,
Chile). Our strategy is to obtain deep images in the $V$ and $R$ 
bands of 21 different fields, covering a total useful area of $\sim 5.1$ 
square degrees. For the spectroscopic confirmation of the SN candidates we 
use FORS1/2 at ESO VLT.
Due to the restriction of time allocation with the VLT only a subsample (the
brightest candidates) could be spectroscopically confirmed.
\section{Data Reduction}
\subsubsection*{Pre-reduction.}
For each search field we take three exposures of 900s with a dithering of few 
arcsec. This allows to remove cosmic rays, detector cosmetic defects and 
moving objects. The pre-reduction, bias and flat fielding, are performed 
using IRAF and MSCRED\footnote{MSCRED is an IRAF package designed to deal
with mosaic images in multi-extension fits (MEF) format. See the manual by
F. Valdes for further details.}. Whenever it was possible, standard sky flats
are improved by constructing a super flat using the science frames. To combine
the dithered frames, these are mapped to a common geometrical grid, properly 
scaled in intensity and finally stacked.
An accurate astrometric calibration (rms $\sim0.3$ arcsec) is attached to the
stacked image using the USNO2 catalogue. The astrometry is used to identify
each object in the fields.
\subsubsection*{Image Subtraction.}
Our SN search technique is based on the subtraction between the image of a
given field and a template image obtained at a previous epoch.
The frame to be searched and the template are first geometrically matched
and trimmed to the overlapping area. To perform the image subtraction we make
use of the ISIS 2.1 package \cite{Alard98}. First, ISIS computes the
convolution kernel to match the image with the best seeing to the other, then
the frames are photometrically scaled and subtracted.
Our experience is that, in any case, the best results are obtained if the two
images have similar seeing.
\subsubsection*{SNe Candidates Selection.}
\begin{figure}[t]
\begin{center}
\includegraphics[width=.47\textwidth]{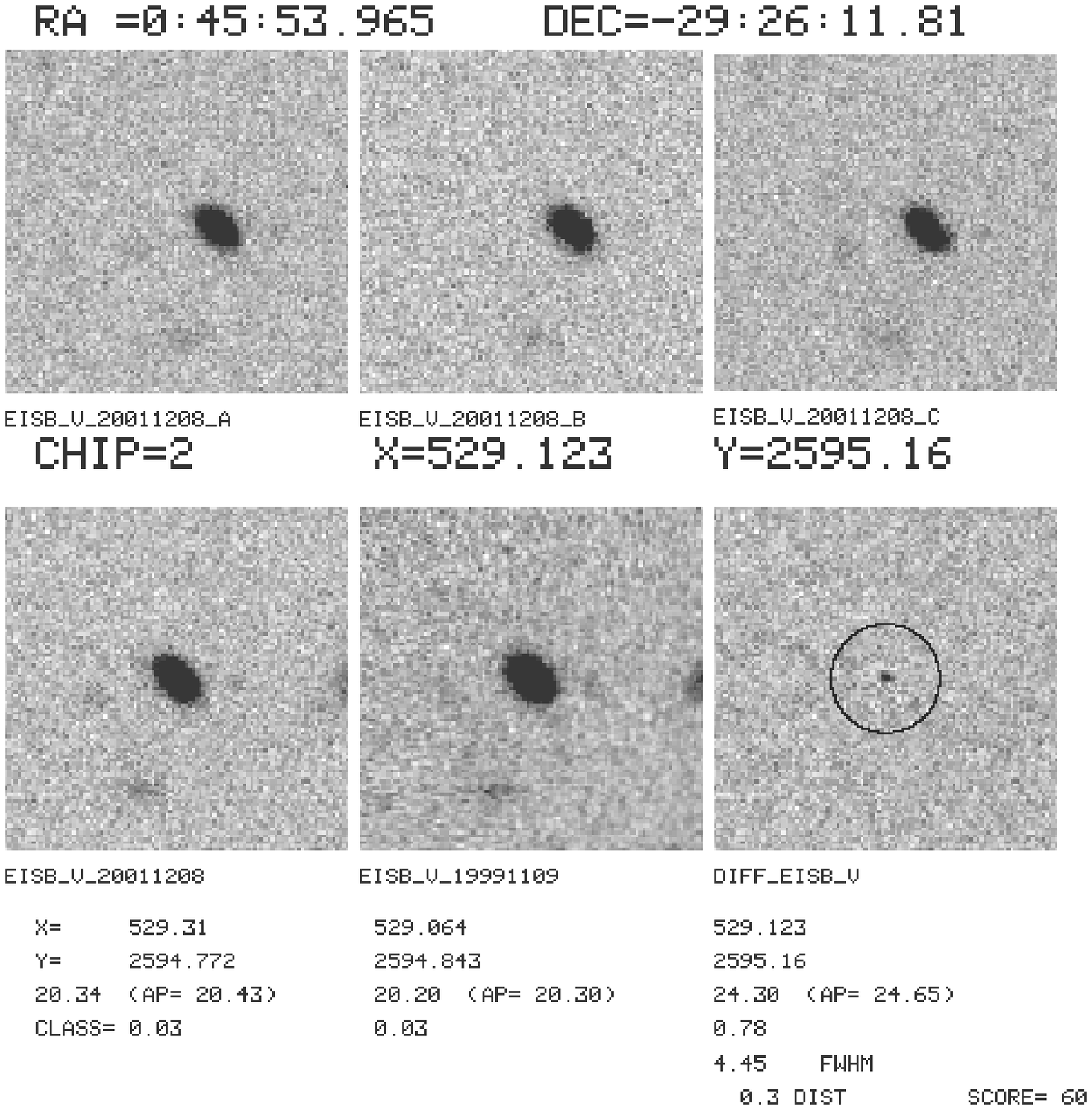}
\includegraphics[width=.47\textwidth]{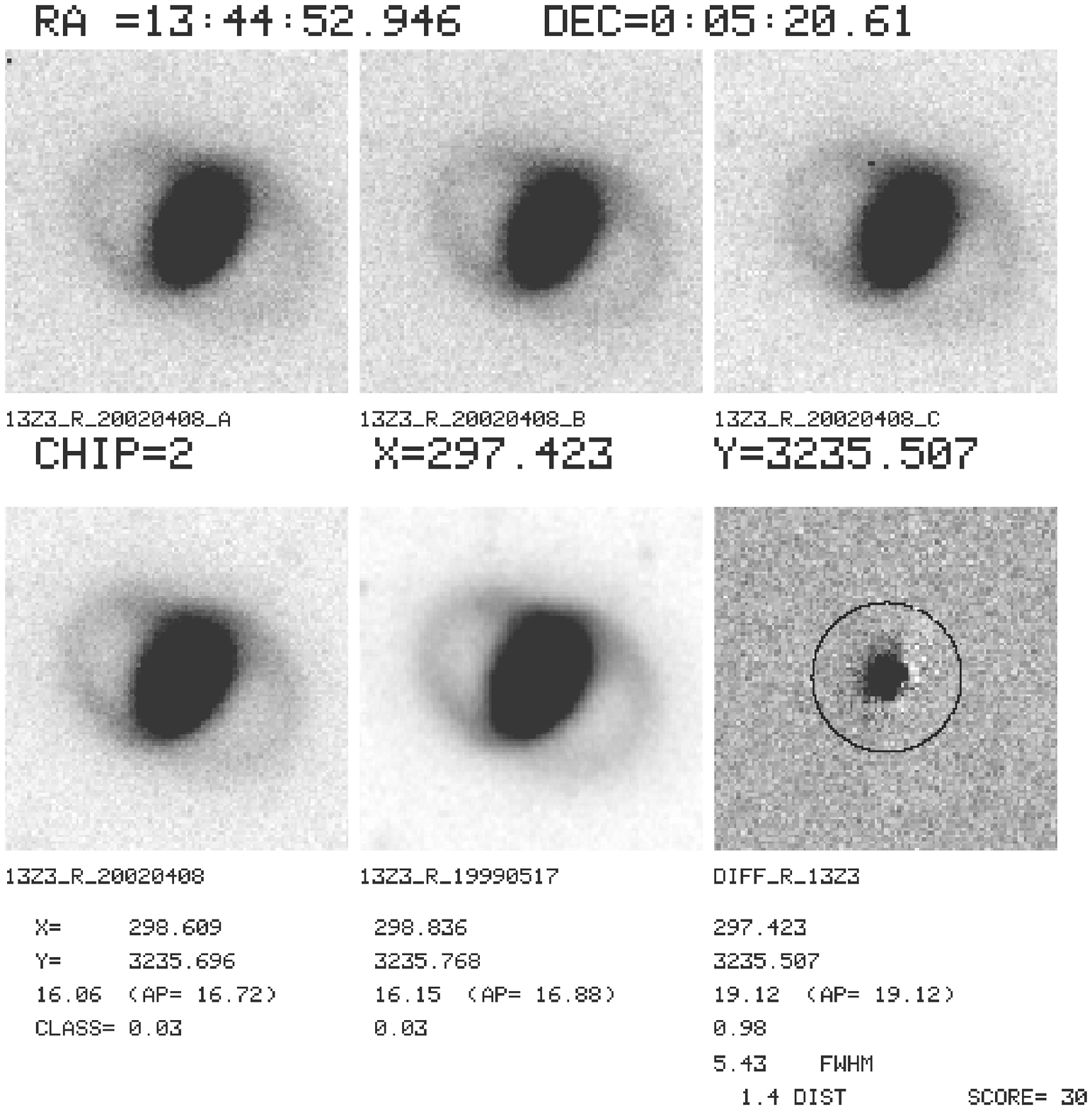}
\end{center}
\caption{Output of the search software. Each image is made of six panels: the 
upper three are the dithering set; at the bottom, from left: the stacked 
image, the reference image and at the the difference image. The panels on the 
left show SN 2001io, one of the 21 spectroscopically confirmed SNe. Those on 
the right show a bright AGN.}
\label{candidate}
\end{figure}

To search the difference image for residuals due to variable sources we use 
Sextractor \cite{sex}. This produces a list which, in general, is heavily 
contaminated by spurious detections due to residuals of bright/saturated 
stars, poorly removed cosmic rays and detector cosmetic defects. In a 
typical field (eight frames) over a thousand detections are usually found. 
To clean as much as possible the spurious detections we have developed an 
automatic procedure which attributes a score to each detection based on 
several parameters (see the next section). After some experience with the WFI 
frames, the scores were calibrated through artificial star experiments. The 
aim is to drastically reduce the number of candidates which are finally 
verified by direct inspection. Eventually, after visual inspection we 
attach a provisional classification (supernovae, active galactic nuclei, 
variable stars and moving objects) to each reliable variable sources 
after considering all the available informations like the stellarity index 
or the distance from the host galaxy nucleus. Figure \ref{candidate} shows
our search software output for a confirmed SN (SN 2001io) and an AGN.
\subsubsection*{Scoring algorithm.}
The scoring procedure takes into account several parameters measured by
Sextractor in the new, template and subtracted images. Saturated objects
produce many spurious residuals in the subtracted image so the first step is 
to purge the list from all detections which are located near bright sources. A 
rank list of the remaining detections is then created according to: the 
stellarity index measured by Sextractor, the FWHM of the object with respect 
to the mean, the distance of the residual with respect to the center of the 
associated object, if any (the host galaxy in the case of SN) and the 
difference between the object magnitudes measured with different 
prescriptions. After some tuning we were able to obtain that the scoring 
algorithm reduces the detection number of a factor 10 without significant 
loss of efficiency in the detection of good SN candidates.
\section{Variable Sources and SN candidates Database.}
Our search technique detects SN candidates but also other variable sources 
like AGN, QSO and variable stars. To reduce such contamination as much as 
possible we exploit the variability history of the fields: the fact that each
of our monitored fields has been observed several times enables us to remove
from the follow--up list the long term variable sources which are not SN.

To record all the detected variable sources and check their variability
history, we developed a MySQL database with a web user interface.
The database contains all informations for each field, the observed epochs and 
the complete list of the variable sources detected. During an observing run, 
with a simple database query from the web interface, it is possible to check if
a given object has already been detected in the past. We found that this tool 
is very effective for the removal of AGNs, which are the only source of
contamination we found. This translates into high efficiency in term of
telescope time spent on targets of interest.
\section{Artificial Star Experiments}
\begin{figure}[t]
\begin{center}
\includegraphics[width=.47\textwidth]{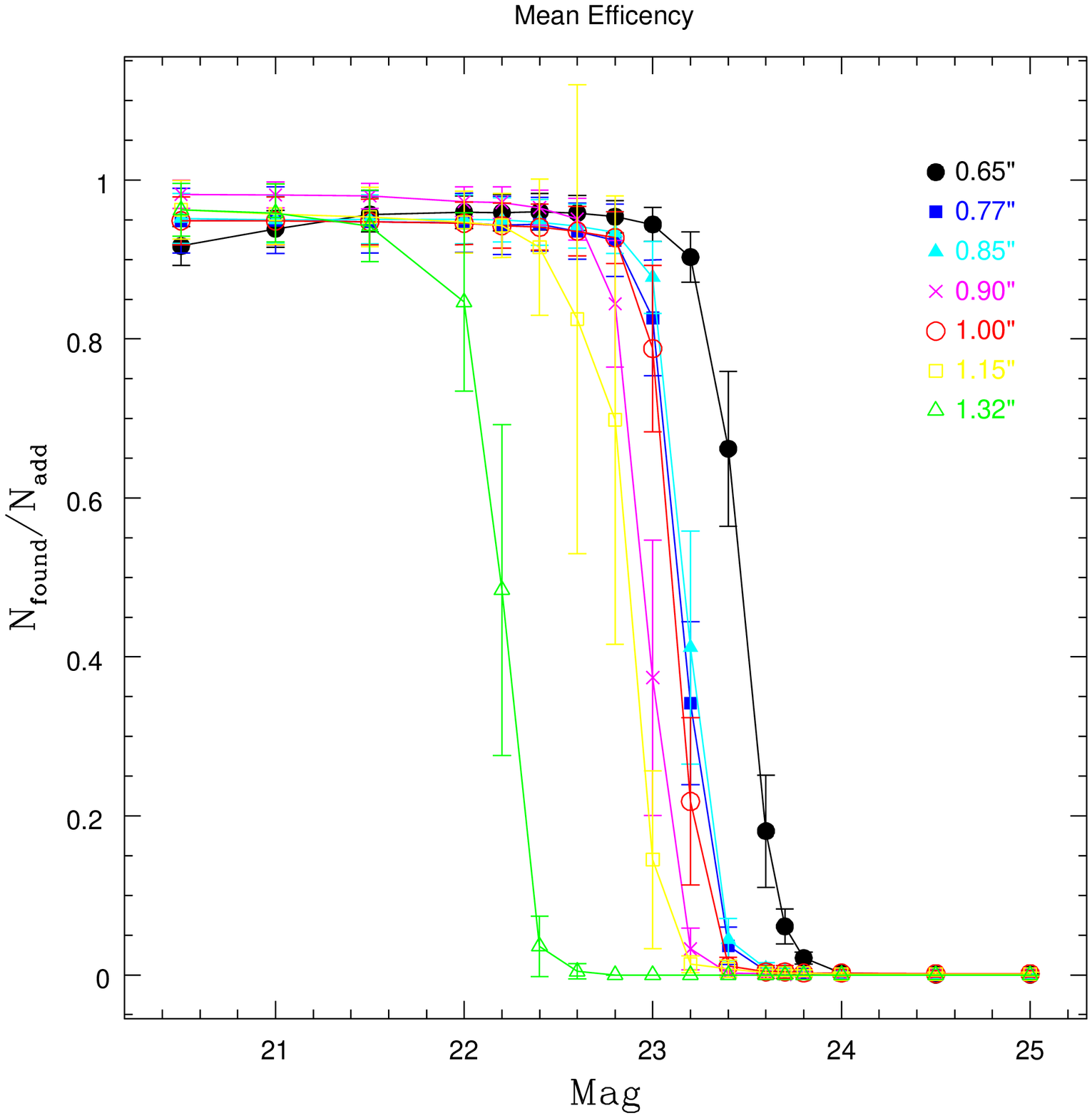}
\includegraphics[width=.47\textwidth]{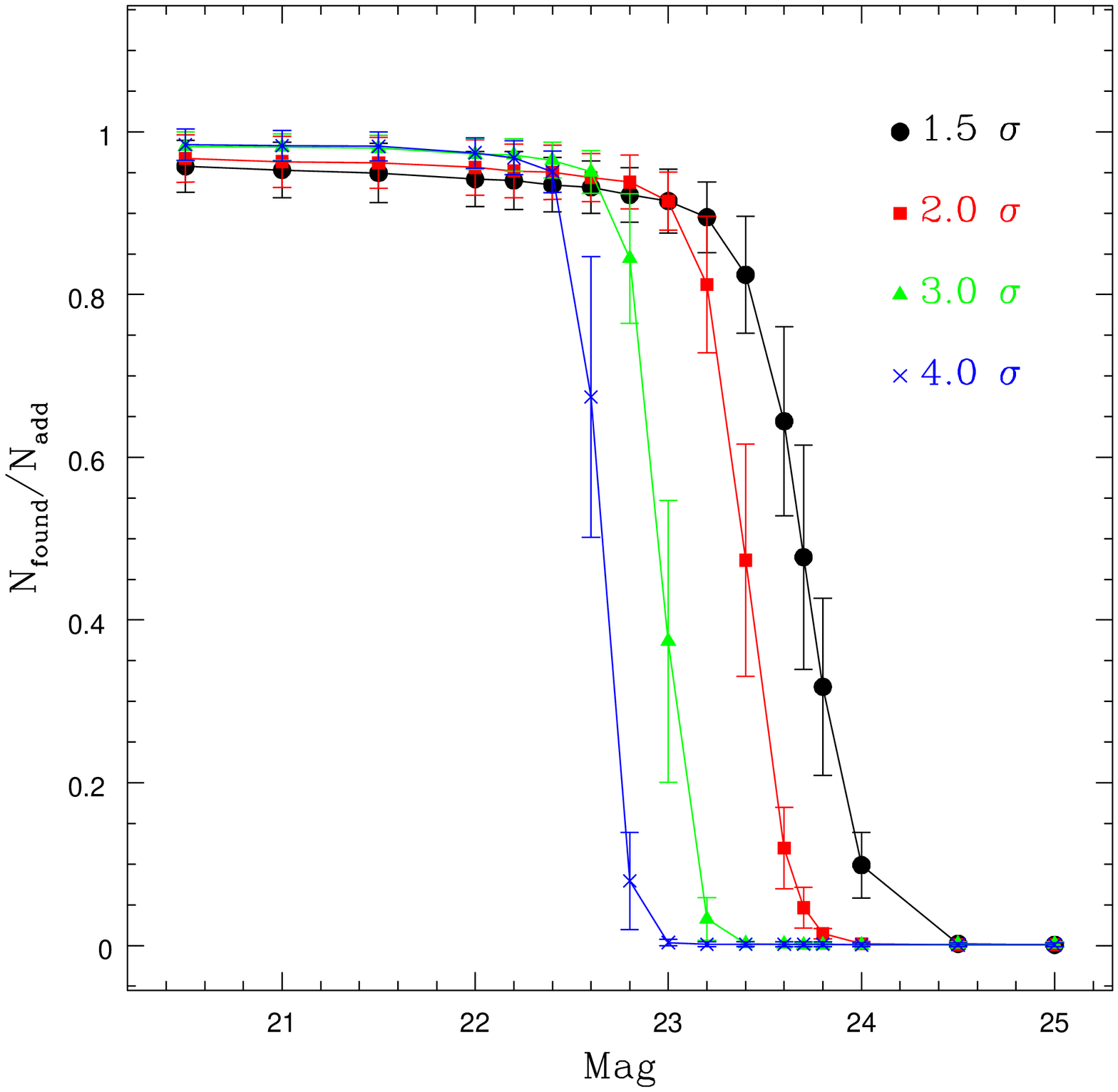}
\caption{\textbf{left}: Dependence of the completeness curves from the image
seeing. \textbf{right}: Dependence of the completeness curves from Sextractor
threshold. The error bars show the dispersion of the efficiency between the
eight frames of the mosaic.}\label{cc}
\end{center}
\end{figure}

In order to obtain an accurate estimate of the search efficiency control time,
which is needed to derive SN rates (cf. by Altavilla et al. in this volume),
we need to construct the completeness curve as a function of magnitude for any
given observation. To build this curve we performed a number of artificial 
star experiments. Artificial stars with the proper PSF and with different 
magnitudes were added to a given image, which is then searched using the same
recipe, as for the real search. These stars were added to each galaxy of the
frame. The distances from the host galaxy nucleus are distributed randomly 
assuming a gaussian distribution with $\sigma$ equal to the FWHM/2.36 of the
galaxy profile.
To test the dependence of the SN detection efficiency from the seeing we ran
several artificial star experiments on images of the same field taken at seven
different epochs and with seeing in the range 0.65--1.32 arcsec (left panel of
Figure \ref{cc}).
For a given seeing, the efficiency drops abruptly from 95$\%$ to 5$\%$ in less
than $\sim0.5$ mag: this helps defining a limiting magnitude for each 
observation because a variation of few tenths of the limiting magnitude has a 
negligible effect on the control time. We define as limiting magnitude the 
point where the detection efficiency is 90\%. It turns out that an improvement
of the seeing of a factor 2 arcsec increases of about 1.5 mag the magnitude 
corresponding to a given efficiency.
Also, in order to check the dependence of the detection efficiency on the 
threshold adopted during the search we ran four experiments on the epoch with
0.90 arcsec seeing using different thresholds (in units of the background rms).
The different detection efficiencies obtained are shown in the right panel of 
Figure \ref{cc}. As expected, the limiting magnitude is fainter for a lower 
threshold, $\sim1$ mag for $1.5\sigma$ compared with $4.0\sigma$. 
However, using a lower threshold has the important drawback of increasing 
significantly the number of spurious detections. Usually a $2.0$--$3.0\sigma$ 
threshold is a reasonable compromise. Further experiments will be performed to
check for variations in efficiency between different fields.
%

%


\begin{thebibliography}{8.}
%
\bibitem{Alard98}  C.~Alard, R.H.~Lupton:  ApJ 
                   \textbf{503}, 325 (1998)
%
\bibitem{sex} E.~Bertin, S.~Arnouts: A\&AS
	      \textbf{117}, 393 (1996)
%
\end{thebibliography}
\end{document}